\pdfoutput=1

\documentclass[11pt,twoside,a4paper,cmspaper,final,collab]{cms-tdr}
\def\svnVersion{248098}\def\cmsCernNo{2013-037}\def\cmsCernDate{\today}\def\cmsMessage{Published in Physics Letters B as \href{http://dx.doi.org/10.1016/j.physletb.2013.12.009}{\doi{10.1016/j.physletb.2013.12.009.}} \\[4ex] \parbox{0.85\textwidth}{ This version includes corrections to be to be published in an upcoming corrigendum that take into account the correct relative uncertainty the measured top-quark pair cross section of 4.1\%. The general conclusions of the paper are unchanged.}}

\begin{document}\cmsNoteHeader{TOP-12-022}

\hyphenation{had-ron-i-za-tion}
\hyphenation{cal-or-i-me-ter}
\hyphenation{de-vices}

\RCS$Revision: 236911 $
\RCS$HeadURL: svn+ssh://alverson@svn.cern.ch/reps/tdr2/papers/TOP-12-022/trunk/TOP-12-022.tex $
\RCS$Id: TOP-12-022.tex 236911 2014-04-16 12:51:09Z snaumann $
\newlength\cmsFigWidth
\ifthenelse{\boolean{cms@external}}{\setlength\cmsFigWidth{0.46\textwidth}}{\setlength\cmsFigWidth{0.75\textwidth}}
\newlength\cmsFigWidthOne\setlength\cmsFigWidthOne{0.46\textwidth}
\ifthenelse{\boolean{cms@external}}{\providecommand{\cmsLeft}{top\xspace}}{\providecommand{\cmsLeft}{left\xspace}}
\ifthenelse{\boolean{cms@external}}{\providecommand{\cmsRight}{bottom\xspace}}{\providecommand{\cmsRight}{right\xspace}}
\newcommand{\mtop}{\ensuremath{m_\cPqt}\xspace}
\newcommand{\mtpole}{\ensuremath{m_\cPqt^{\scriptscriptstyle{\text{pole}}}}\xspace}
\newcommand{\mZ}{\ensuremath{m_\cPZ}\xspace}
\cmsNoteHeader{TOP-12-022} 
\title{Determination of the top-quark pole mass and strong coupling constant from the \ttbar production cross section in pp collisions at $\sqrt{s} = 7\TeV$}

\date{\today}

\abstract{
  The inclusive cross section for top-quark pair production measured by the CMS experiment in proton-proton collisions at a
  center-of-mass energy of 7\TeV is compared to the QCD prediction at next-to-next-to-leading order with various
  parton distribution functions to determine the top-quark pole mass, \mtpole,
  or the strong coupling constant, $\alpha_S$.
  With the parton distribution function set NNPDF2.3, a pole mass of $176.7{}^{+3.0}_{-2.8}$\GeV is obtained
  when constraining $\alpha_S$ at the scale of the Z boson mass, \mZ, to the current world average.
  Alternatively, by constraining
  \mtpole to the latest average from direct mass measurements,
  a value of $\alpha_S (\mZ) = 0.1151 {}^{+0.0028}_{-0.0027}$
  is extracted. This is the first determination of $\alpha_S$ using events from top-quark production.
}

\hypersetup{%
pdfauthor={CMS Collaboration},%
pdftitle={Determination of the top-quark pole mass and strong coupling constant from the ttbar production cross section in pp collisions at sqrt(s) = 7 TeV},%
pdfsubject={CMS},%
pdfkeywords={CMS, physics, top, quark, pair, cross section, mass, QCD, strong, coupling, constant}}

\maketitle 

\section{Introduction}
\label{sec:intro}

The Large Hadron Collider (LHC) has provided a wealth of proton-proton collisions,
which has enabled the Compact Muon Solenoid (CMS) experiment \cite{Chatrchyan:2008zzk}
to measure cross sections for the production of top-quark pairs (\ttbar)
with high precision employing a variety of approaches \cite{Chatrchyan:2011nb,Chatrchyan:2011ew,Chatrchyan:2011yy,Chatrchyan:2012vs,Chatrchyan:2012bra,Chatrchyan:2012ria,Chatrchyan:2013kff,Chatrchyan:2013ual,Chatrchyan:2012saa}.
Comparing the presently available results, obtained at a center-of-mass energy, $\sqrt{s}$, of 7\TeV,
to theoretical predictions allows for stringent tests of the underlying models
and for constraints on fundamental parameters.
Top-quark pair production can be described in the framework of quantum chromodynamics (QCD) and
calculations for the inclusive \ttbar cross section, $\sigma_{\ttbar}$,
have recently become available to complete next-to-next-to-leading order (NNLO)
in perturbation theory \cite{Czakon:2013goa}.
Crucial inputs to these calculations are: the top-quark mass, $\mtop$; the strong coupling constant, $\alpha_S$;
and the gluon distribution in the proton,
since \ttbar production at LHC energies is expected to occur predominantly via gluon-gluon fusion.

The top-quark mass is one of the fundamental parameters of the standard model (SM) of particle physics.
Its value significantly affects predictions for many observables either directly or
via radiative corrections.
As a consequence, the measured $\mtop$ is one of the key inputs to electroweak precision fits,
which enable comparisons between experimental results and predictions within and beyond the SM.
Furthermore, together with the Higgs-boson mass and $\alpha_S$, $\mtop$ has direct implications on the stability of the
electroweak vacuum \cite{Degrassi:2012ry,Alekhin:2012py}.
The most precise result for $\mtop$, obtained by combining direct measurements performed at the Tevatron,
is $173.18 \pm 0.94$\GeV \cite{Aaltonen:2012ra}.
Similar measurements performed by the CMS Collaboration \cite{Chatrchyan:2011nb,Chatrchyan:2012cz,Chatrchyan:2012ea,Chatrchyan:2013boa}
are in agreement with the Tevatron result and of comparable precision.
However, except for a few cases \cite{Chatrchyan:2013boa},
these direct measurements rely on the relation between $\mtop$ and the respective experimental observable,
\eg, a reconstructed invariant mass, as expected from simulated events.
In QCD beyond leading order, $\mtop$ depends on the renormalization scheme \cite{Hoang:2008xm,Ahrens:2011px}.
The available Monte Carlo generators contain matrix elements at leading order or next-to-leading order (NLO), while
higher orders are simulated by applying parton showering.
Studies suggest that $\mtop$ as implemented in Monte Carlo generators corresponds approximately
to the pole (``on-shell'') mass, $\mtpole$, but that the value of the true pole mass could be of the order of 1\GeV higher
compared to $\mtop$ in the current event generators \cite{Buckley:2011ms}.
In addition to direct $\mtop$ measurements,
the mass dependence of the QCD prediction for $\sigma_{\ttbar}$ can be used to determine $\mtop$
by comparing the measured to the predicted cross section \cite{Langenfeld:2009wd,Abazov:2011pta,Ahrens:2011px,Beneke:2011mq,Beneke:2012wb,Alekhin:2012py}.
Although the sensitivity of $\sigma_{\ttbar}$ to $\mtop$ might not be strong enough
to make this approach competitive in precision,
it yields results affected by different sources of systematic uncertainties compared
to the direct $\mtop$ measurements and allows for extractions of $\mtop$ in theoretically well-defined mass schemes.
It has been advocated to directly extract the $\overline{\textrm{MS}}$ mass of the top quark
using the $\sigma_{\ttbar}$ prediction in that scheme \cite{Langenfeld:2009wd}.
The relation between pole and $\overline{\textrm{MS}}$ mass is known to three-loop level in QCD
but might receive large electroweak corrections \cite{Jegerlehner:2012kn}.
In principle,
the difference between the results obtained when extracting $\mtop$ in the pole and
converting it to the $\overline{\textrm{MS}}$ scheme or extracting the $\overline{\textrm{MS}}$ mass directly
should be small in view of the precision that the extraction of $\mtop$ from the inclusive $\sigma_{\ttbar}$
at a hadron collider provides.
Therefore, only the pole mass scheme is employed in this Letter.

With the exception of the quark masses, $\alpha_S$ is the only free parameter of the QCD Lagrangian. While the
renormalization group equation predicts the energy dependence of the strong coupling, \ie,
gives a functional form for $\alpha_S (Q)$, where $Q$ is the energy scale of the process,
actual values of $\alpha_S$ can only be obtained based on experimental data.
By convention and to facilitate comparisons, $\alpha_S$ values measured at different energy scales
are typically evolved to $Q = \mZ$, the mass of the Z boson.
The current world average for $\alpha_S (\mZ)$ is $0.1184 \pm 0.0007$ \cite{PhysRevD.86.010001}.
In spite of this relatively precise result,
the uncertainty on $\alpha_S$ still contributes significantly to many QCD predictions, including expected cross sections
for top-quark pairs or Higgs bosons.
Furthermore, thus far very few measurements allow $\alpha_S$ to be tested at high $Q$
and the precision on the average for $\alpha_S (\mZ)$ is driven by low-$Q$ measurements.
Energies up to 209\GeV were probed with hadronic final states in electron-positron collisions
at LEP using NNLO predictions \cite{Dissertori:2009ik,OPAL:2011aa,Dissertori:2009qa,Abbate:2010xh}.
Jet measurements at the Tevatron and the LHC have recently
extended the range up to 400\GeV \cite{:2012xib}, 600\GeV \cite{Malaescu:2012ts}, and 1.4\TeV~\cite{Chatrchyan:2013txa}.
However, most predictions for jet production in hadron collisions are only available up to NLO QCD.
Even when these predictions are available at
approximate NNLO, as used in \cite{Abazov:2009nc},
they suffer from significant uncertainties related to the choice and variation of the renormalization and
factorization scales, $\mu_R$ and $\mu_F$, as well as from uncertainties related to non-perturbative corrections.

In cross section calculations, $\alpha_s$ appears not only in
the expression for the parton-parton interaction but also in the QCD evolution of the
parton distribution functions (PDFs).
Varying the value of $\alpha_S(\mZ)$ in the $\sigma_{\ttbar}$ calculation therefore
requires a consistent modification of the PDFs.
Moreover, a strong correlation between $\alpha_S$ and the gluon PDF at large partonic momentum fractions
is expected to significantly enhance the sensitivity of $\sigma_{\ttbar}$ to $\alpha_S$ \cite{Czakon:2013tha}.

In this Letter, the predicted $\sigma_{\ttbar}$ is compared to the most precise single measurement to
date~\cite{Chatrchyan:2012bra}, and values of $\mtpole$ and $\alpha_S (\mZ)$ are determined.
This extraction is performed under the assumption that the
measured $\sigma_{\ttbar}$ is not affected by non-SM physics.
The interplay of the values of $\mtpole$, $\alpha_s$ and the proton PDFs
in the prediction of $\sigma_{\ttbar}$ is studied.
Five different PDF sets, available at NNLO, are employed and for each a series of
different choices of $\alpha_S(\mZ)$ are considered.
A simultaneous extraction of top-quark mass and strong coupling constant
from the total \ttbar cross section alone is not possible
since both parameters alter the predicted $\sigma_{\ttbar}$ in such a way that
any variation of one parameter can be compensated by a variation of the other.
Values of $\mtpole$ and $\alpha_S(\mZ)$
are therefore determined at fixed values of $\alpha_S (\mZ)$ and $\mtpole$, respectively.
For the $\mtpole$ extraction, $\alpha_S (\mZ)$ is constrained to the latest world average value
with its corresponding uncertainty ($0.1184 \pm 0.0007$) \cite{PhysRevD.86.010001}.
Furthermore, it is assumed that the $\mtop$ parameter of the Monte Carlo generator that is employed in the
$\sigma_{\ttbar}$ measurement is equal to $\mtpole$ within $\pm 1.00$\GeV \cite{Buckley:2011ms}.
For the $\alpha_S$ extraction, $\mtpole$ is set to the Tevatron average of
$173.18 \pm 0.94$\GeV \cite{Aaltonen:2012ra}.
To account for the possible difference between the pole mass and the Monte Carlo generator mass \cite{Buckley:2011ms},
an additional uncertainty, assumed to be 1.00\GeV, is added in quadrature to the experimental uncertainty,
resulting in a total uncertainty on the top-quark mass constraint, $\delta \mtpole$, of 1.4\GeV.
Although the potential $\alpha_S$ dependence of the direct $\mtop$ measurements has not been explicitly evaluated,
it is assumed to be covered by the quoted mass uncertainty.

\section{Predicted Cross Section}
\label{sec:prediction}

The expected $\sigma_{\ttbar}$ has been calculated to NNLO for all production channels,
namely the all-fermionic scattering modes ($\Pq\Paq$, $\Pq\Pq'$, $\Pq\Paq'$, $\Pq\Pq$ $\to \ttbar + X$)
\cite{Baernreuther:2012ws,Czakon:2012zr}, the reaction $\Pq\Pg \to \ttbar + X$ \cite{Czakon:2012pz},
and the dominant process $\Pg\Pg \to \ttbar + X$ \cite{Czakon:2013goa}.
In the present analysis, these calculations are used as implemented in the program \textsc{Top++}~2.0~\cite{Czakon:2011xx}.
Soft-gluon resummation is performed at next-to-next-leading-log (NNLL) accuracy
\cite{Beneke:2009rj,Czakon:2009zw}.
The scales $\mu_R$ and $\mu_F$ are set to $\mtpole$.
In order to evaluate the theoretical uncertainty of the fixed-order calculation, the missing contributions
from higher orders are estimated by varying $\mu_R$ and $\mu_F$ up and down by a factor of 2 independently,
while using the restriction $0.5 \leq \mu_F / \mu_R \leq 2$.
These choices for the central scale and the variation procedure
were suggested by the authors of the NNLO calculations
and used for earlier $\sigma_{\ttbar}$ predictions as well \cite{Cacciari:2008zb}.

Five different NNLO PDF sets are employed: ABM11~\cite{Alekhin:2012ig}, CT10~\cite{Gao:2013xoa},
HERAPDF1.5~\cite{HERA15NNLO}, MSTW\-2008~\cite{Martin:2009iq,Martin:2009bu}, and NNPDF2.3~\cite{Ball:2012cx}.
The corresponding uncertainties are calculated at the 68\% confidence level for all PDF sets.
This is done by recalculating the $\sigma_{\ttbar}$ at NNLO+NNLL for each of the provided
eigenvectors or replicas of the respective PDF set and then performing error propagation
according to the prescription of that PDF group.
In the specific case of the CT10 PDF set, the uncertainties are provided for the 90\% confidence
level only. For this Letter, following the recommendation of the CTEQ group, these uncertainties
are adjusted using the general relation between confidence intervals based on Gaussian distributions
\cite{PhysRevD.86.010001}, \ie, scaled down by a factor of $\sqrt{2}  \erf^{-1}(0.90) = 1.64$,
where $\erf$ denotes the error function.

The dependence of the predicted $\sigma_{\ttbar}$ on the choice of $\mtpole$ is studied by
varying $\mtpole$ in the range from 130 to 220\GeV in steps of 1\GeV
and found to be well described by a third-order polynomial in $\mtpole$ divided by $(\mtpole)^4$.
The $\alpha_S$ dependence of $\sigma_{\ttbar}$ is studied by varying the
value of $\alpha_S (\mZ)$ over the entire valid range for a particular PDF set, as listed
in Table~\ref{tab:PDFalphaScans}.
The relative change of $\sigma_{\ttbar}$ as a function of $\alpha_S(\mZ)$ can be
parametrized using a second-order polynomial in $\alpha_S(\mZ)$,
where the three coefficients of that polynomial depend linearly on $\mtpole$.

\begin{table*}[htb]
  \begin{center}
    \topcaption{Default $\alpha_S (\mZ)$ values and $\alpha_S (\mZ)$ variation ranges of the NNLO PDF sets
      used in this analysis. Because the NNPDF2.3 PDF set does not have a default value of $\alpha_S (\mZ)$,
  preferring to provide the full uncertainties and systematic variations for various $\alpha_S (\mZ)$ points,
  the $\alpha_S (\mZ)$ value obtained by the NNPDF Collaboration
  with NNPDF2.1 \cite{Ball:2011us} is used.
      The step size for the $\alpha_S (\mZ)$ scans is 0.0010 in all cases.
      The uncertainties on the default values are shown for illustration purposes only.
      \label{tab:PDFalphaScans}}
    \begin{tabular}{lcccc}
      &                    &               & \multicolumn{2}{c}{Provided $\alpha_S (\mZ)$ scan} \\
      & Default $\alpha_S (\mZ)$ & Uncertainty & Range & \# of points \\
      \hline
      ABM11      & 0.1134 & $\pm$ 0.0011 & 0.1040--0.1200 & 17 \\
      CT10       & 0.1180 & $\pm$ 0.0020 & 0.1100--0.1300 & 21 \\
      HERAPDF1.5 & 0.1176 & $\pm$ 0.0020 & 0.1140--0.1220 & 9  \\
      MSTW2008   & 0.1171 & $\pm$ 0.0014 & 0.1070--0.1270 & 21 \\
      NNPDF2.3   & 0.1174 & $\pm$ 0.0007 & 0.1140--0.1240 & 11 \\
    \end{tabular}
  \end{center}
\end{table*}

\begin{figure}[htb]
  \centering
  \includegraphics[width=\cmsFigWidthOne]{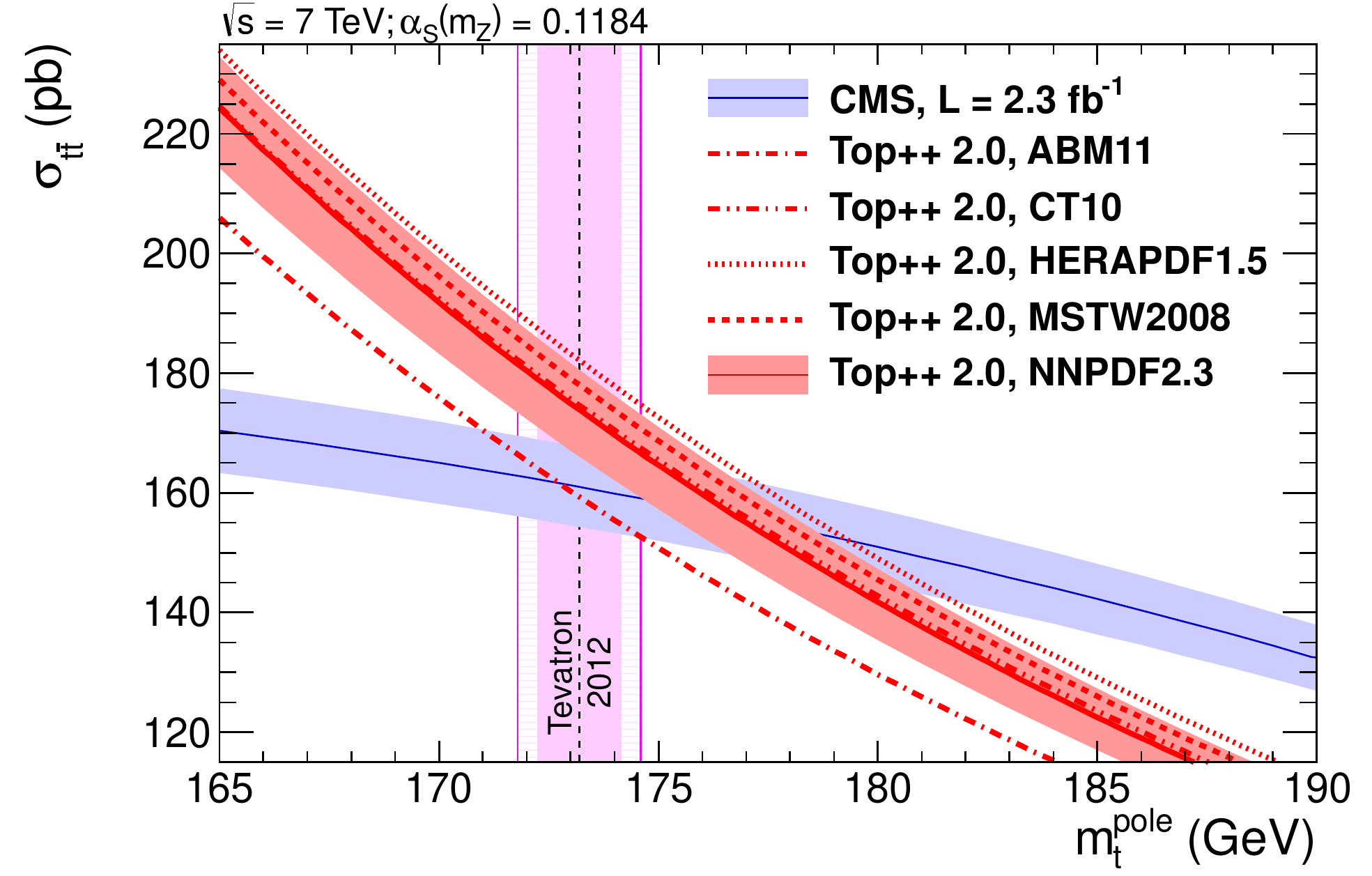} \hfill
  \includegraphics[width=\cmsFigWidthOne]{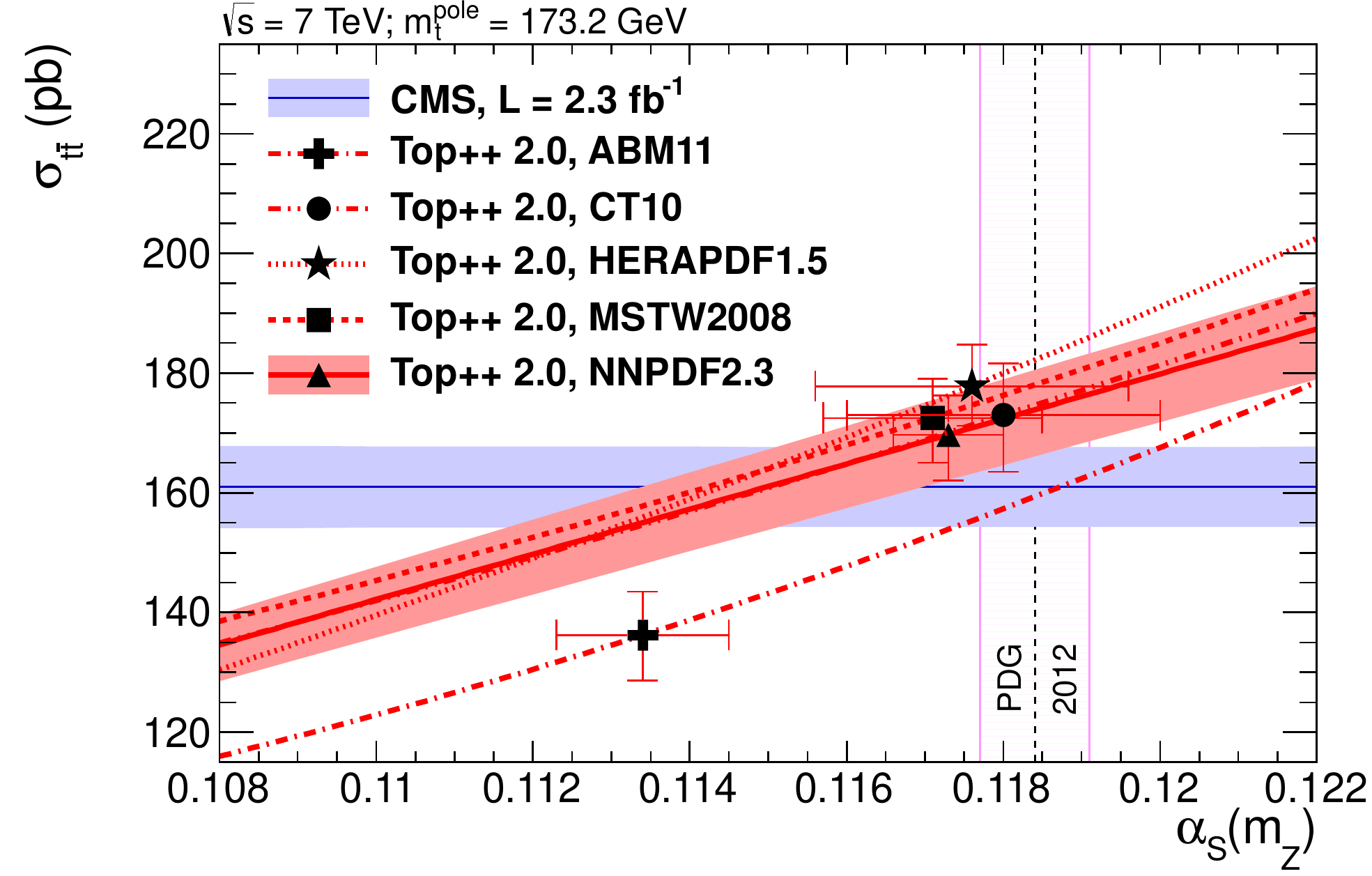}
  \caption{Predicted \ttbar cross section at NNLO+NNLL, as a function of the top-quark pole mass (\cmsLeft)
    and of the strong coupling constant (\cmsRight),
    using five different NNLO PDF sets, compared to the cross section measured by CMS
    assuming $\mtop = \mtpole$.
    The uncertainties on the measured $\sigma_{\ttbar}$
    as well as the renormalization and factorization scale and PDF uncertainties
    on the prediction with NNPDF2.3 are illustrated with filled bands.
    The uncertainties on the $\sigma_{\ttbar}$ predictions using the other PDF sets are indicated
    only in the \cmsRight panel at the corresponding default $\alpha_S (\mZ)$ values.
    The $\mtpole$ and $\alpha_S (\mZ)$ regions favored
    by the direct measurements at the Tevatron and by the latest world average, respectively,
    are shown as hatched areas.
    In the \cmsLeft panel, the inner (solid) area of the vertical band corresponds to the
    original uncertainty of the direct $\mtop$ average, while the outer (hatched) area
    additionally accounts for the possible difference between this mass and $\mtpole$.
    \label{fig:xsec_vs_alpha}}
\end{figure}

The resulting $\sigma_{\ttbar}$ predictions are compared in Fig.~\ref{fig:xsec_vs_alpha}, both as a function
of $\mtpole$ and of $\alpha_S (\mZ)$.
For a given value of $\alpha_S (\mZ)$, the predictions based on NNPDF2.3 and CT10
are very similar. The cross sections obtained with MSTW2008 and HERAPDF1.5 are slightly higher
while the predictions obtained with ABM11 are significantly lower due to a smaller gluon density
in the relevant kinematic range \cite{Alekhin:2012ig}.
In addition to the absolute normalization, differences in the slope of $\sigma_{\ttbar}$ as a function
of $\alpha_S (\mZ)$ are observed between some of the PDF sets.

\section{Measured Cross Section}

In this Letter, the most precise single measurement for $\sigma_{\ttbar}$ \cite{Chatrchyan:2012bra} is used.
It was derived at $\sqrt{s}$ = 7\TeV by the CMS Collaboration
from data collected in 2011 in the dileptonic decay channel and corresponding to an integrated luminosity of 2.3\fbinv.
Assuming $\mtop = 172.5$\GeV and $\alpha_S (\mZ) = 0.1180$, the observed cross section is $161.9\pm6.7\unit{pb}$.
Systematic effects on this measurement from the choice and uncertainties of the PDFs
were studied and found to be negligible.

The measured $\sigma_{\ttbar}$ shows a dependence on the value of $\mtop$ that is used in the Monte Carlo simulations
since the change in the event kinematics affects the expected selection efficiency
and thus the acceptance corrections that are employed to infer $\sigma_{\ttbar}$ from the observed event yield.
A parametrization for this dependence, which is illustrated in Fig.~\ref{fig:xsec_vs_alpha},
was already given in Section~8 of Ref.~\cite{Chatrchyan:2012bra}.
At $\mtop = 173.2$\GeV, for example, the observed cross section is 161.0\unit{pb}.
The relative uncertainty of 4.1\% on the measured $\sigma_{\ttbar}$ is independent of $\mtop$ to very good approximation.

Changes of the assumed value of $\alpha_S (\mZ)$ in the simulation used to derive the acceptance corrections
can alter the measured $\sigma_{\ttbar}$ as well, which is discussed in this Letter for the first time.
QCD radiation effects increase at higher $\alpha_S (\mZ)$,
both at the matrix-element level and at the hadronization level. The $\alpha_S (\mZ)$-dependence of the
acceptance corrections
is studied using the NLO CTEQ6AB PDF sets \cite{Pumplin:2005rh},
and the \textsc{Powheg Box}~1.4 \cite{Alioli:2010xd,Frixione:2007nw}
NLO generator for \ttbar production
interfaced with \PYTHIA~6.4.24 \cite{Sjostrand:2006za} for the parton showering.
Additionally, the impact of $\alpha_S (\mZ)$ variations on the acceptance
is studied with standalone \PYTHIA as a plain leading-order generator with parton showering
and cross-checked with \MCFM~6.2 \cite{Campbell:2012uf} as an NLO prediction without parton showering.
In all cases, a relative change of the acceptance by
less than $1\%$ is observed when varying $\alpha_S (\mZ)$ by $\pm 0.0100$ with respect to the CTEQ
reference value of 0.1180.
This is accounted for by applying an $\alpha_S (\mZ)$-dependent uncertainty to the measured
$\sigma_{\ttbar}$.
This additional uncertainty is also
included in the uncertainty band shown in Fig.~\ref{fig:xsec_vs_alpha}.
Over the relevant $\alpha_S (\mZ)$ range,
there is almost no increase in the total uncertainty of 4.1\% on the measured $\sigma_{\ttbar}$.

In the $\mtop$ and $\alpha_S (\mZ)$ regions favored
by the direct measurements at the Tevatron and by the latest world average, respectively,
the measured and the predicted cross section are compatible within their uncertainties
for all considered PDF sets.
When using ABM11 with its default $\alpha_S (\mZ)$, the discrepancy between
measured and predicted cross section is larger than one standard deviation.

\section{Probabilistic Approach}

In the following, the theory prediction for $\sigma_{\ttbar}$ is employed to construct a Bayesian prior
to the cross section measurement, from which a joint posterior in $\sigma_{\ttbar}$, $\mtpole$ and $\alpha_S (\mZ)$
is derived.
Finally, this posterior is marginalized by integration over $\sigma_{\ttbar}$ and a Bayesian confidence interval
for $\mtpole$ or $\alpha_S (\mZ)$ is computed based on the external constraint for $\alpha_S (\mZ)$
or $\mtpole$, respectively.

\begin{figure}[htb]
  \centering
  \includegraphics[width=\cmsFigWidthOne]{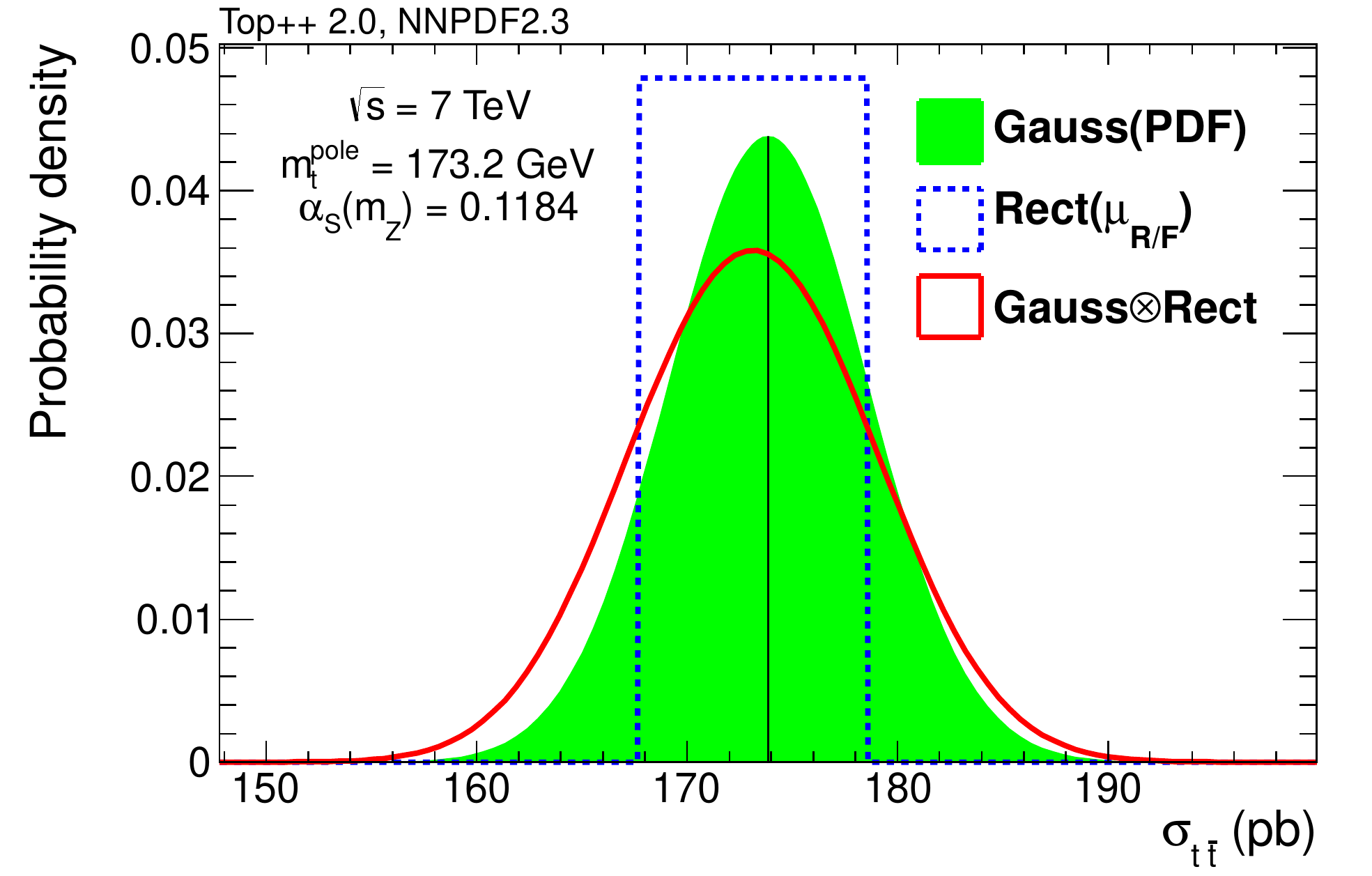}
  \caption{Probability distributions for the predicted \ttbar cross section at NNLO+NNLL
    with $\mtpole = 173.2$\GeV, $\alpha_S (\mZ) = 0.1184$ and the NNLO parton distributions from NNPDF2.3.
    The resulting probability, $f_{\text{th}} (\sigma_{\ttbar})$, represented by a solid line,
    is obtained by convolving a Gaussian distribution (filled area) that accounts for the PDF uncertainty
    with a rectangular function (dashed line) that covers the scale variation uncertainty.}
  \label{fig:convolution}
\end{figure}

The probability function for the predicted cross section,
$f_{\text{th}} (\sigma_{\ttbar})$, is obtained through an analytic convolution of two probability distributions,
one accounting for the PDF uncertainty and the other for scale uncertainties.
A Gaussian distribution of width $\delta_{\text{PDF}}$ is used to describe the PDF uncertainty.
Given that no particular probability distribution is known that should be adequate
for the confidence interval obtained from the variation of $\mu_R$ and $\mu_F$ \cite{Cacciari:2008zb},
the corresponding uncertainty on the $\sigma_{\ttbar}$ prediction is approximated using a flat prior, \ie, a rectangular
function that provides equal probability over the whole range covered by the scale variation and vanishes elsewhere.
The resulting probability function is given by:
\begin{linenomath} \begin{equation*}
  f_{\textrm{th}} (\sigma_{\ttbar}) =
  \frac{1}{2 \left( \sigma_{\ttbar}^{(h)} - \sigma_{\ttbar}^{(l)} \right)}
  \left( \erf \left[ \frac{\sigma_{\ttbar}^{(h)} - \sigma_{\ttbar}}{\sqrt{2}\, \delta_{\text{PDF}}} \right] -
    \erf \left[  \frac{\sigma_{\ttbar}^{(l)} - \sigma_{\ttbar}}{\sqrt{2}\, \delta_{\text{PDF}}} \right] \right).
\end{equation*} \end{linenomath}
Here, $\sigma_{\ttbar}^{(l)}$ and $\sigma_{\ttbar}^{(h)}$ denote the lowest and the highest cross section values,
respectively, that are obtained when varying $\mu_R$ and $\mu_F$ as described in Section~\ref{sec:prediction}.
An example for the resulting probability distributions is shown in Fig.~\ref{fig:convolution}.

\begin{figure}[htb]
  \centering
  \includegraphics[width=\cmsFigWidthOne]{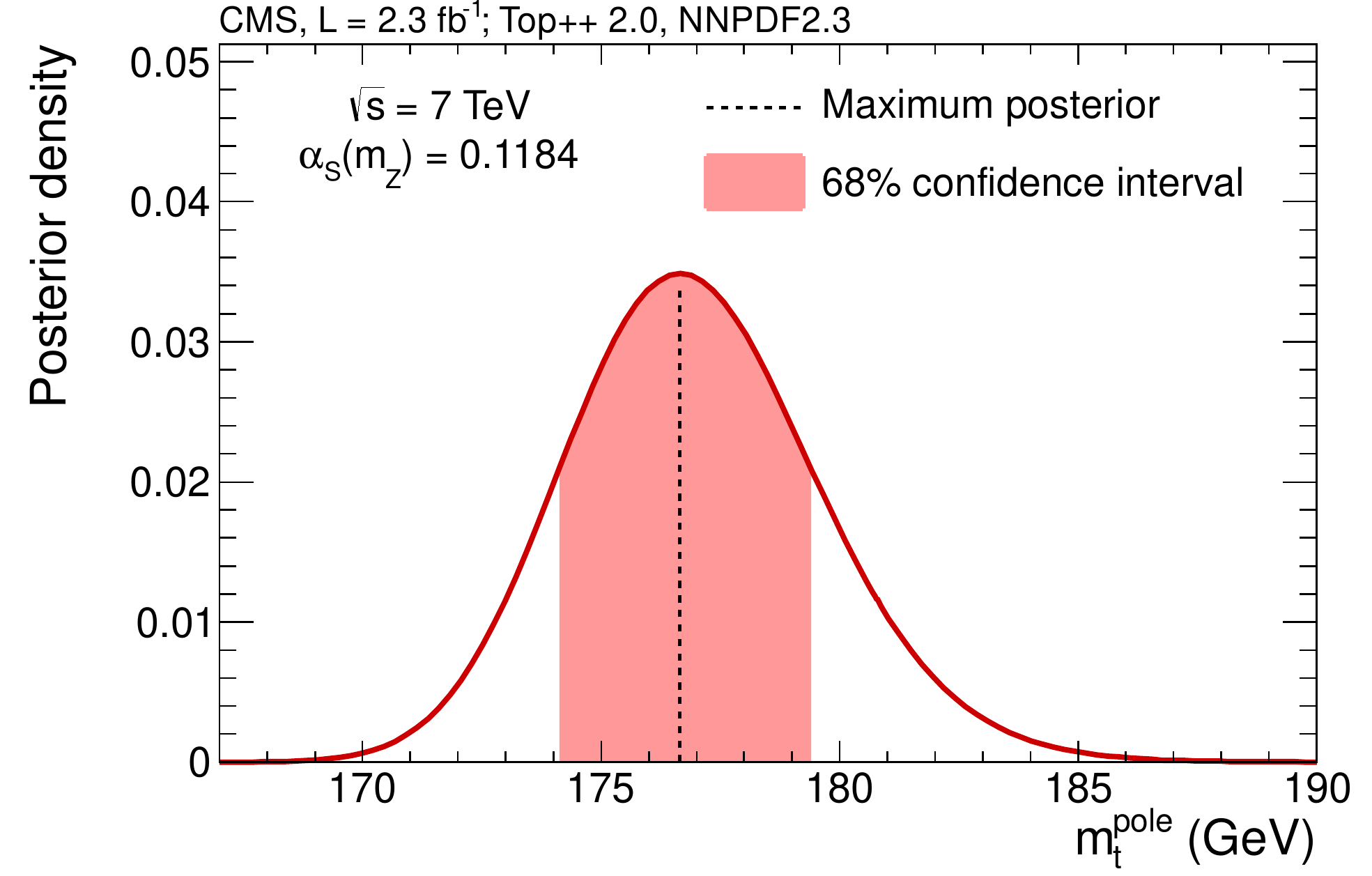} \hfill
  \includegraphics[width=\cmsFigWidthOne]{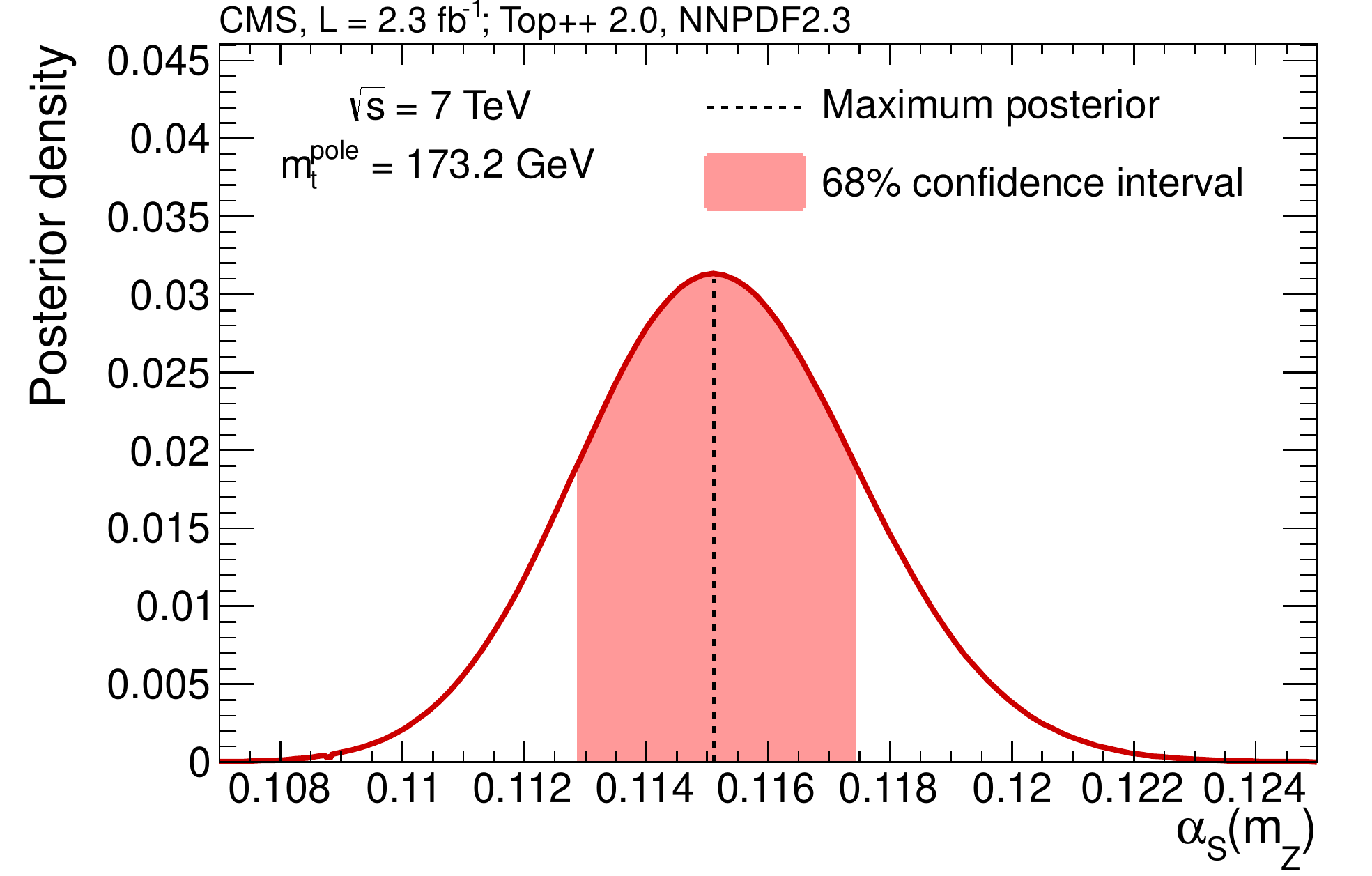}
  \caption{Marginal posteriors $P (\mtpole)$ (\cmsLeft) and $P (\alpha_S)$ (\cmsRight)
    based on the cross section prediction at NNLO+NNLL with the
    NNLO parton distributions from NNPDF2.3.
    The posteriors are constructed as described in the text.
    Here, $P (\mtpole)$ is shown for $\alpha_S (\mZ) = 0.1184$ and $P (\alpha_S)$ for
    \mbox{$\mtpole = 173.2$\GeV}.}
  \label{fig:finalLikelihoods}
\end{figure}

The probability distribution $f_{\text{th}} (\sigma_{\ttbar})$ is multiplied by another Gaussian
probability, $f_{\text{exp}} (\sigma_{\ttbar})$, which represents the measured cross section and its uncertainty,
to obtain the most probable $\mtpole$ or $\alpha_S (\mZ)$ value for a
given $\alpha_S (\mZ)$ or $\mtpole$, respectively, from the maximum of the marginalized posterior:
\begin{linenomath} \begin{equation*}
  P(x) = \int f_{\text{exp}} (\sigma_{\ttbar} | x) \ f_{\text{th}} (\sigma_{\ttbar} | x) \,\rd\sigma_{\ttbar},
  \quad x = \mtpole,\; \alpha_S (\mZ).
\end{equation*} \end{linenomath}
Examples of $P (\mtpole)$ and $P (\alpha_S)$ are shown in Fig.~\ref{fig:finalLikelihoods}. Confidence intervals are
determined from the 68\% area around the maximum of the posterior and requiring equal function values at
the left and right edges.

The approximate contributions of the uncertainties on the measured and the predicted cross sections to the width
of this Bayesian confidence interval can be estimated by repeatedly rescaling the size of the corresponding uncertainty
component. The widths of the obtained confidence intervals are then used to extrapolate to the case in which a given
component vanishes.

To assess the impact of the uncertainties on the $\alpha_S (\mZ)$ and $\mtpole$ values that are used as constraints
in the present analysis, $P (\mtpole)$ is re-evaluated at $\alpha_S (\mZ)$ = 0.1177 and 0.1191,
reflecting the $\pm$0.0007 uncertainty on the $\alpha_S (\mZ)$ world average,
and $P (\alpha_S)$ is re-evaluated at $\mtpole$ = 171.8 and 174.6\GeV,
reflecting the $\delta \mtpole$ = 1.4\GeV as explained in Section~\ref{sec:intro}.
The resulting shifts in the most likely values of $\mtpole$ and $\alpha_S (\mZ)$
are added in quadrature to those obtained
from the 68\% areas of the posteriors calculated with the central values of the constraints.

\section{Results and Conclusions}

\begin{figure}[thbp]
  \centering
  \includegraphics[width=\cmsFigWidth]{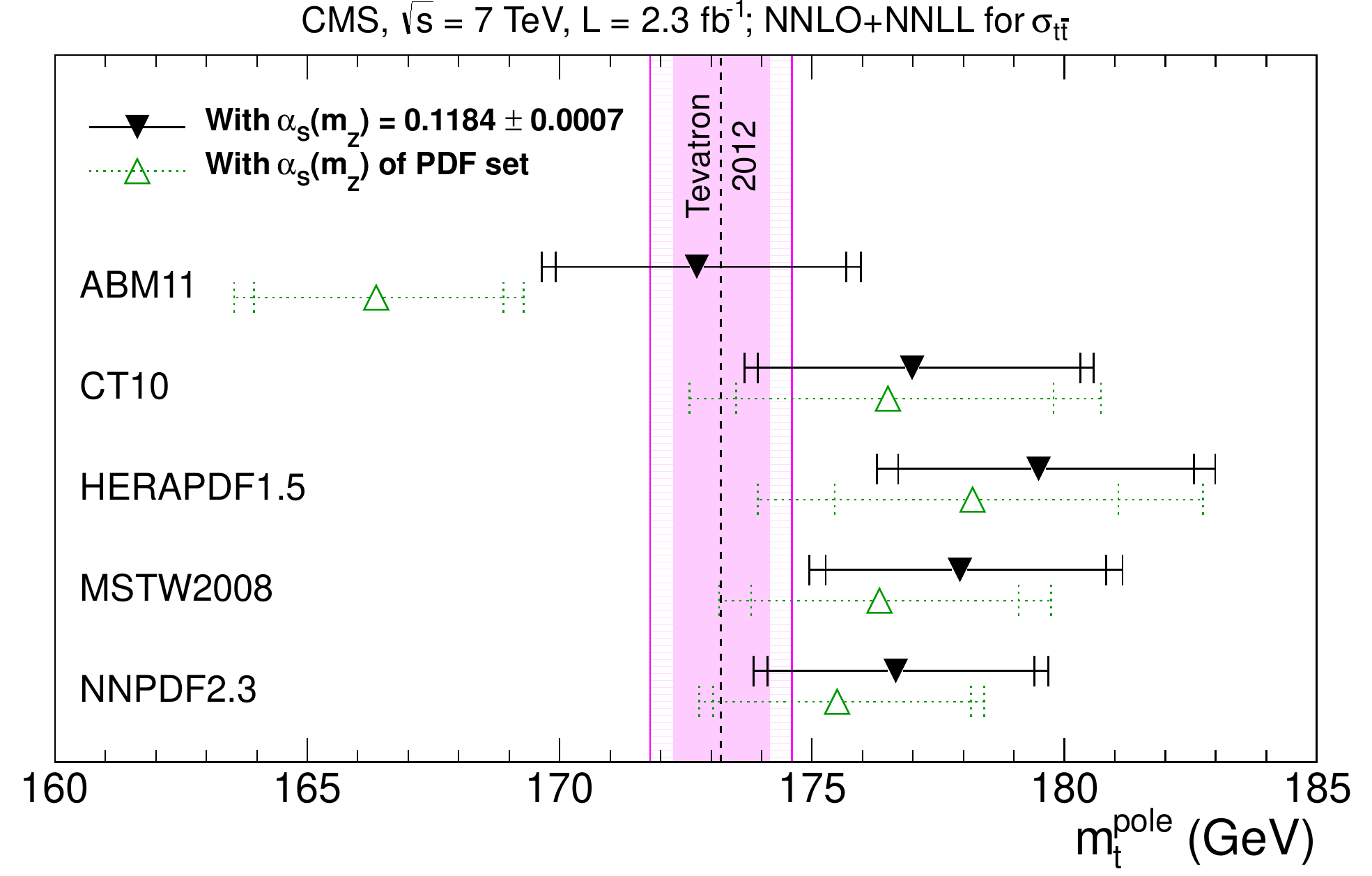}
  \caption{Results obtained for $\mtpole$ from
    the measured \ttbar cross section together with the prediction at NNLO+NNLL using different NNLO PDF sets.
    The filled symbols represent the results obtained when using the $\alpha_S (\mZ)$ world average,
    while the open symbols indicate the results obtained with the default $\alpha_S (\mZ)$ value
    of the respective PDF set.
    The inner error bars include the uncertainties on the measured cross section and on
    the LHC beam energy as well as the PDF and scale uncertainties on the predicted cross section.
    The outer error bars additionally account for the uncertainty on the $\alpha_S (\mZ)$ value used
    for a specific prediction.
    For comparison, the latest average of direct $\mtop$ measurements
    is shown as vertical band, where
    the inner (solid) area corresponds to the
    original uncertainty of the direct $\mtop$ average, while the outer (hatched) area
    additionally accounts for the possible difference between this mass and $\mtpole$.
  \label{fig:massSummaryPlot}}
\end{figure}

Values of the top-quark pole mass determined using the \ttbar cross section measured by CMS together
with the cross section prediction from NNLO+NNLL QCD and five different NNLO PDF sets
are listed in Table~\ref{tab:resultsMass}.
These values are extracted under the assumption that the $\mtop$
parameter in the Monte Carlo generator that was employed to obtain the mass-dependent
acceptance correction of the measured cross section, shown in Fig.~\ref{fig:xsec_vs_alpha},
is equal to the pole mass.
A difference of 1.0\GeV between the two mass definitions \cite{Buckley:2011ms} would
result in changes of 0.3--0.6\GeV in the extracted pole masses, which is included as a systematic uncertainty.
As illustrated in Fig.~\ref{fig:massSummaryPlot}, the results based on NNPDF2.3, CT10, MSTW2008, and HERAPDF1.5
are higher than the latest average of direct $\mtop$ measurements
but generally compatible within the uncertainties.
They are also consistent with the indirect determination of the top-quark pole mass
obtained in the electroweak fits \cite{Eberhardt:2012gv,Baak:2012kk}
when employing the mass of the new boson discovered at the LHC \cite{Aad:2012tfa,Chatrchyan:2012ufa}
under the assumption that this is the SM Higgs boson.
The central $\mtpole$ value obtained with the ABM11 PDF set,
which has a significantly smaller gluon density than the other PDF sets,
is also compatible with the average from direct $\mtop$ measurements.
Note, however, that all these results in Table~\ref{tab:resultsMass}
are obtained employing the $\alpha_S (\mZ)$ world average of $0.1184 \pm 0.0007$,
while ABM11 with its default $\alpha_S (\mZ)$ of $0.1134 \pm 0.0011$ would yield an $\mtpole$ value of $166.4^{+2.9}_{-2.8}$\GeV.
\begin{table*}[pthb]
  \begin{center}
    \renewcommand{\arraystretch}{1.25}
    \topcaption{Results obtained for $\mtpole$ by comparing the measured \ttbar cross section to
      the NNLO+NNLL prediction with different NNLO PDF sets.
      The total uncertainties account for
      the full uncertainty on the measured cross section ($\sigma_{\ttbar}^{\text{meas}}$),
      the PDF and scale ($\mu_{R,F}$) uncertainties on the predicted cross section,
      the uncertainties of the $\alpha_S(\mZ)$ world average
      and of the LHC beam energy ($E_{\text{LHC}}$), and
      the ambiguity in translating the MC-generator based mass dependence ($\mtop^{\text{MC}}$) of the measured cross section
      into the pole-mass scheme.
     \label{tab:resultsMass}}
\begin{tabular}{lcccccccc}
 & \multirow{2}{*}{$\mtpole$ (\GeVns{})} & \multicolumn{7}{c}{Uncertainty on $\mtpole$ (\GeVns{})} \\ \cline{3-9}
 & & Total & $\sigma_{\ttbar}^{\text{meas}}$ & PDF & $\mu_{R,F}$ & $\alpha_{S}$ & $E_{\text{LHC}}$ & $\mtop^{\text{MC}}$ \\
\hline
ABM11      &  172.7 & ${}^{+3.2}_{-3.1}$ & ${}^{+1.8}_{-1.8}$ & ${}^{+2.2}_{-2.0}$ & ${}^{+0.7}_{-0.7}$ & ${}^{+1.0}_{-1.0}$ & ${}^{+0.8}_{-0.8}$ & ${}^{+0.4}_{-0.3}$\\
CT10       &  177.0 & ${}^{+3.6}_{-3.3}$ & ${}^{+2.2}_{-2.1}$ & ${}^{+2.4}_{-2.0}$ & ${}^{+0.9}_{-0.9}$ & ${}^{+0.8}_{-0.8}$ & ${}^{+0.9}_{-0.9}$ & ${}^{+0.5}_{-0.4}$ \\
HERAPDF1.5 &  179.5 & ${}^{+3.5}_{-3.2}$ & ${}^{+2.4}_{-2.2}$ & ${}^{+1.7}_{-1.5}$ & ${}^{+0.9}_{-0.8}$ & ${}^{+1.2}_{-1.1}$ & ${}^{+1.0}_{-1.0}$ & ${}^{+0.6}_{-0.5}$ \\
MSTW2008   &  177.9 & ${}^{+3.2}_{-3.0}$ & ${}^{+2.2}_{-2.1}$ & ${}^{+1.6}_{-1.4}$ & ${}^{+0.9}_{-0.9}$ & ${}^{+0.9}_{-0.9}$ & ${}^{+0.9}_{-0.9}$ & ${}^{+0.5}_{-0.5}$ \\
NNPDF2.3   &  176.7 & ${}^{+3.0}_{-2.8}$ & ${}^{+2.1}_{-2.0}$ & ${}^{+1.5}_{-1.3}$ & ${}^{+0.9}_{-0.9}$ & ${}^{+0.7}_{-0.7}$ & ${}^{+0.9}_{-0.9}$ & ${}^{+0.5}_{-0.4}$ \\
\end{tabular}
  \end{center}
\end{table*}

\begin{figure}[htbp]
  \centering
  \includegraphics[width=\cmsFigWidth]{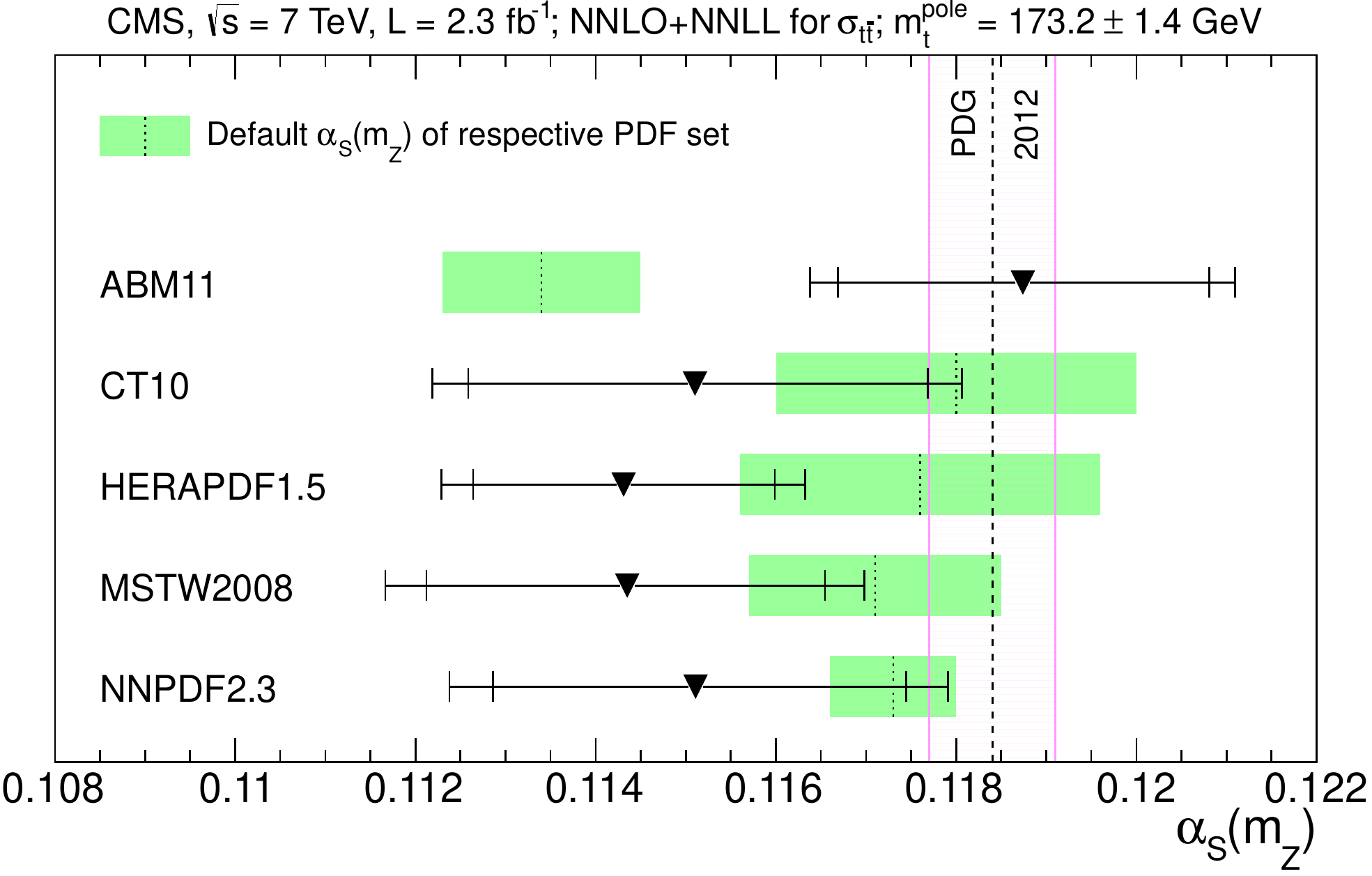}
  \caption{Results obtained for $\alpha_S (\mZ)$ from
    the measured \ttbar cross section together with the prediction at NNLO+NNLL using different NNLO PDF sets.
    The inner error bars include the uncertainties on the measured cross section and on
    the LHC beam energy as well as the PDF and scale uncertainties on the predicted cross section.
    The outer error bars additionally account for the uncertainty on $\mtpole$.
    For comparison, the latest $\alpha_S (\mZ)$ world average with its uncertainty is shown as a hatched band.
    For each PDF set, the default $\alpha_S (\mZ)$ value and its uncertainty are indicated using a dotted line and a
    shaded band.
    \label{fig:alphaSummaryPlot}}
\end{figure}

The $\alpha_S (\mZ)$ values obtained when fixing the value of $\mtpole$ to $173.2 \pm 1.4\GeV$,
\ie, inverting the logic of the extraction, are listed in Table~\ref{tab:resultsAlpha}.
As illustrated in Fig.~\ref{fig:alphaSummaryPlot},
the results obtained using NNPDF2.3, CT10, MSTW2008, and HERAPDF1.5 are lower than the
$\alpha_S (\mZ)$ world average but in most cases still compatible with it within the uncertainties.
While the $\alpha_S (\mZ)$ value obtained with ABM11 is compatible with the world average,
it is significantly different from the default $\alpha_S (\mZ)$ of this PDF set.
\begin{table*}[htb]
  \begin{center}
    \renewcommand{\arraystretch}{1.25}
    \topcaption{Results obtained for $\alpha_S (\mZ)$ by comparing the measured \ttbar cross section to
      the NNLO+NNLL prediction with different NNLO PDF sets.
      The total uncertainties account for
      the full uncertainty on the measured cross section ($\sigma_{\ttbar}^{\text{meas}}$),
      the PDF and scale ($\mu_{R,F}$) uncertainties on the predicted cross section,
      the uncertainty assigned to the knowledge of $\mtpole$,
      and the uncertainty of the LHC beam energy ($E_{\text{LHC}}$).
      \label{tab:resultsAlpha}}

\begin{tabular}{lccccccc}
 & \multirow{2}{*}{$\alpha_S (\mZ)$} & \multicolumn{6}{c}{Uncertainty on $\alpha_S (\mZ)$} \\ \cline{3-8}
 & & Total & $\sigma_{\ttbar}^{\text{meas}}$ & PDF & $\mu_{R,F}$ & $\mtpole$ & $E_{\text{LHC}}$ \\
\hline
ABM11      & 0.1187 & ${}^{+0.0024}_{-0.0024}$ & ${}^{+0.0013}_{-0.0015}$ & ${}^{+0.0015}_{-0.0014}$ & ${}^{+0.0006}_{-0.0005}$ & ${}^{+0.0010}_{-0.0010}$ & ${}^{+0.0006}_{-0.0006}$\\
CT10       & 0.1151 & ${}^{+0.0030}_{-0.0029}$ & ${}^{+0.0018}_{-0.0018}$ & ${}^{+0.0018}_{-0.0016}$ & ${}^{+0.0008}_{-0.0007}$ & ${}^{+0.0012}_{-0.0013}$ & ${}^{+0.0007}_{-0.0007}$\\
HERAPDF1.5 & 0.1143 & ${}^{+0.0020}_{-0.0020}$ & ${}^{+0.0012}_{-0.0013}$ & ${}^{+0.0010}_{-0.0009}$ & ${}^{+0.0005}_{-0.0004}$ & ${}^{+0.0010}_{-0.0010}$ & ${}^{+0.0006}_{-0.0006}$\\
MSTW2008   & 0.1144 & ${}^{+0.0026}_{-0.0027}$ & ${}^{+0.0017}_{-0.0018}$ & ${}^{+0.0012}_{-0.0011}$ & ${}^{+0.0008}_{-0.0007}$ & ${}^{+0.0012}_{-0.0013}$ & ${}^{+0.0007}_{-0.0008}$\\
NNPDF2.3   & 0.1151 & ${}^{+0.0028}_{-0.0027}$ & ${}^{+0.0017}_{-0.0018}$ & ${}^{+0.0013}_{-0.0011}$ & ${}^{+0.0009}_{-0.0008}$ & ${}^{+0.0013}_{-0.0013}$ & ${}^{+0.0008}_{-0.0008}$\\
\end{tabular}
  \end{center}
\end{table*}

Modeling the uncertainty related to the choice and variation of the renormalization and factorization scales
with a Gaussian instead of the flat prior results in only minor changes of the
$\mtpole$ and $\alpha_S (\mZ)$ values and uncertainties.
With the precise NNLO+NNLL calculation,
these scale uncertainties are found to be of the size of 0.7--0.9\GeV on $\mtpole$
and 0.0004--0.0009 on $\alpha_S (\mZ)$,
\ie, of the order of 0.3--0.8\%.

The energy of the LHC beams is known to an accuracy of 0.65\%~\cite{Wenninger:1546734}
and thus the center-of-mass energy of 7\TeV with an uncertainty of $\pm$46\GeV.
Based on the expected dependence of $\sigma_{\ttbar}$ on $\sqrt{s}$,
this can be translated into an additional uncertainty of $\pm$1.8\% on the comparison of the measured to the
predicted \ttbar cross section,
which yields an additional uncertainty of $\pm$(0.5--0.7)\% on the obtained $\mtpole$ and
$\alpha_S (\mZ)$ values.

For the main results of this Letter,
the $\mtpole$ and $\alpha_S (\mZ)$ values determined with the parton densities of NN\-PDF2.3 are used.
The primary motivation is that parton distributions derived using the NNPDF methodology can be explicitly shown
to be parametrization independent, in the sense that results are unchanged even when the number of input
parameters is substantially increased \cite{Ball:2008by}.

In summary, a top-quark pole mass of $176.7{}^{+3.0}_{-2.8}$\GeV is obtained by comparing the measured cross section
for inclusive \ttbar production in proton-proton collisions at $\sqrt{s}$ = 7\TeV to QCD calculations at NNLO+NNLL.
Due to the small uncertainty on the measured cross section and the state-of-the-art NNLO calculations,
the precision of this result is higher compared to earlier determinations of $\mtpole$ following the same approach.
This extraction provides an important test of the mass scheme applied in Monte Carlo simulations
and gives complementary information, with different sensitivity to theoretical and experimental uncertainties,
than direct measurements of $\mtop$.
Alternatively, $\alpha_S (\mZ) = 0.1151 {}^{+0.0028}_{-0.0027}$ is obtained from the \ttbar cross section
when constraining $\mtpole$ to 173.2 $\pm$1.4\GeV.
This is the first determination of the strong coupling constant from top-quark production
and the first $\alpha_S (\mZ)$ result at full NNLO QCD obtained at a hadron collider.

\section*{Acknowledgements}

We thank Alexander Mitov for his help with the NNLO calculations.
We congratulate our colleagues in the CERN accelerator departments for the excellent performance of the LHC and thank the technical and administrative staffs at CERN and at other CMS institutes for their contributions to the success of the CMS effort. In addition, we gratefully acknowledge the computing centres and personnel of the Worldwide LHC Computing Grid for delivering so effectively the computing infrastructure essential to our analyses. Finally, we acknowledge the enduring support for the construction and operation of the LHC and the CMS detector provided by the following funding agencies: BMWF and FWF (Austria); FNRS and FWO (Belgium); CNPq, CAPES, FAPERJ, and FAPESP (Brazil); MES (Bulgaria); CERN; CAS, MoST, and NSFC (China); COLCIENCIAS (Colombia); MSES (Croatia); RPF (Cyprus); MoER, SF0690030s09 and ERDF (Estonia); Academy of Finland, MEC, and HIP (Finland); CEA and CNRS/IN2P3 (France); BMBF, DFG, and HGF (Germany); GSRT (Greece); OTKA and NKTH (Hungary); DAE and DST (India); IPM (Iran); SFI (Ireland); INFN (Italy); NRF and WCU (Republic of Korea); LAS (Lithuania); CINVESTAV, CONACYT, SEP, and UASLP-FAI (Mexico); MSI (New Zealand); PAEC (Pakistan); MSHE and NSC (Poland); FCT (Portugal); JINR (Dubna); MON, RosAtom, RAS and RFBR (Russia); MESTD (Serbia); SEIDI and CPAN (Spain); Swiss Funding Agencies (Switzerland); NSC (Taipei); ThEPCenter, IPST, STAR and NSTDA (Thailand); TUBITAK and TAEK (Turkey); NASU (Ukraine); STFC (United Kingdom); DOE and NSF (USA).

\bibliography{auto_generated}   

\cleardoublepage \appendix\section{The CMS Collaboration \label{app:collab}}\begin{sloppypar}\hyphenpenalty=5000\widowpenalty=500\clubpenalty=5000\textbf{Yerevan Physics Institute,  Yerevan,  Armenia}\\*[0pt]
S.~Chatrchyan, V.~Khachatryan, A.M.~Sirunyan, A.~Tumasyan
\vskip\cmsinstskip
\textbf{Institut f\"{u}r Hochenergiephysik der OeAW,  Wien,  Austria}\\*[0pt]
W.~Adam, T.~Bergauer, M.~Dragicevic, J.~Er\"{o}, C.~Fabjan\cmsAuthorMark{1}, M.~Friedl, R.~Fr\"{u}hwirth\cmsAuthorMark{1}, V.M.~Ghete, N.~H\"{o}rmann, J.~Hrubec, M.~Jeitler\cmsAuthorMark{1}, W.~Kiesenhofer, V.~Kn\"{u}nz, M.~Krammer\cmsAuthorMark{1}, I.~Kr\"{a}tschmer, D.~Liko, I.~Mikulec, D.~Rabady\cmsAuthorMark{2}, B.~Rahbaran, C.~Rohringer, H.~Rohringer, R.~Sch\"{o}fbeck, J.~Strauss, A.~Taurok, W.~Treberer-Treberspurg, W.~Waltenberger, C.-E.~Wulz\cmsAuthorMark{1}
% ...
\vskip\cmsinstskip
% ...
1:~~Also at Vienna University of Technology, Vienna, Austria\\
2:~~Also at CERN, European Organization for Nuclear Research, Geneva, Switzerland\\
% ...

\end{sloppypar}
\end{document}